# The influence of magnetic-moment relaxation motion on second viscosity in superfluid solutions

**N.I. Pushkina**
*Scientific Research Computation Centre, Moscow State University, Vorob'yovy Gory, Moscow, 119992 Russia*
(E-mail: N.Pushkina@mererand.com)

**The influence of $He^3$ nuclear magnetization relaxation on the second viscosity in quantum solutions subjected to an external oscillating magnetic field is studied. The cases of first –, second– and forth–sound waves are examined. The expressions for the second viscosity coefficients are derived and the conditions for the viscosity to be decreased are analyzed.**

The second viscosity in liquids and gases is known to reveal itself if the volume, i.e. the density, of a medium is changed. If in a medium there take place fairly slow relaxation processes with a considerable relaxation time the thermodynamic equilibrium may not be attained along with the volume changes and the process will then be thermodynamically irreversible. This results in energy absorption associated with the second viscosity. Thus, the value of the second viscosity depends on relative values of the relaxation time and the characteristic times of motions that can take place in a medium, such as e.g. the propagation of sound waves. The investigation of relaxation process influence on the second viscosity and the general method of approaching the problem are given in Ref. [1]. The method is also described in the book [2]. If a medium is subjected to an external means of overpopulating internal energy levels, relaxation processes in the medium might, in principle, not only decrease the second viscosity, but even render it negative due to an external energy source. In classical media the peculiarities of sound-wave propagation and amplification in the presence of external sources affecting the population of internal energy levels and through this the second viscosity were considered previously in literature, see e.g. Refs. [3, 4, 5]. In Ref. [4] an electric discharge as an external energy source induced the excitation of vibrational degrees of freedom in a molecular gas. In [5] it was shown that the anomalous behaviour of acoustic waves in such media is associated with the inversion of the second viscosity, that is with the change of its sign. In the present paper the influence of the relaxation process associated with $He^3$ nuclear magnetization on the second viscosity in $He^3$– $He^4$ superfluid solutions subjected to an external oscillating magnetic field is studied. The theory is developed for both the first- and the second-sound waves and also for the case of sound propagation in narrow capillaries, that is for the fourth-sound wave. Up to now similar phenomena have not been investigated in quantum liquids.

The equations of superfluid solution dynamics with only the second viscosity retained are of the form [6]

$$\frac{\partial \rho}{\partial t} + \text{div } \mathbf{j} = 0,$$

$$\frac{\partial j_i}{\partial t} + \frac{\partial \Pi_{ik}}{\partial x_k} = \frac{\partial}{\partial x_i}\left[\xi_1 \text{ div }(\mathbf{j} - \rho \mathbf{v}_n) + \xi_2 \text{ div } \mathbf{v}_n\right],$$

$$\frac{\partial \mathbf{v}_s}{\partial t} + \nabla\left(\mu - c\frac{z}{\rho}\right) = \nabla\left[\xi_3 \text{ div }(\mathbf{j} - \rho \mathbf{v}_n) + \xi_4 \text{ div } \mathbf{v}_n\right], \quad (1)$$

$$\frac{\partial (\rho c)}{\partial t} + \text{div }(\rho c \mathbf{v}_n) = 0,$$

$$\frac{\partial (\rho \sigma)}{\partial t} + \text{div }(\rho \sigma \mathbf{v}_n) = 0.$$

In these equations $\rho$ is the density of the solution; $\mathbf{j} = \rho_s \mathbf{v}_s + \rho_n \mathbf{v}_n$ is the mass flux density; $\rho_s, \mathbf{v}_s$ and $\rho_n, \mathbf{v}_n$ are superfluid and normal densities and velocities; $\Pi_{ik} = p\delta_{ik} + \rho_s v_{si} v_{sk} + \rho_n v_{ni} v_{nk}$ is the momentum flux density tensor; $c$ is the concentration of $He^3$ in the solution; $\mu$ is the chemical potential per unit mass; $z$ is the thermodynamic potential; $\sigma$ is the entropy; $\xi_1, \xi_2, \xi_3, \xi_4$ are the second viscosity coefficients. For the problem under consideration it is enough to limit oneself with the linear approximation.

We shall describe the tendency of the magnetic system to come to thermal equilibrium with its surroundings in the presence of an external magnetic field by an equation of the form

$$\frac{\partial m}{\partial t} = -\frac{1}{\tau}(m - m_0) + \chi \frac{\partial H}{\partial t}. \quad (2)$$

Here $m_0$ is the equilibrium magnetization value; $\tau$ is the relaxation time; $\chi$ is the magnetic susceptibility; $H$ is the external magnetic field. At sound propagation the equilibrium magnetization value changes together with thermodynamic quantities and $m_0$ can be presented as $m_0 = m_{00} + m'_0$, where $m_{00}$ is the steady-state value of $m_0$ corresponding to average thermodynamic quantities, while $m'_0$ is the periodic part varying as $e^{-i\omega t}$. Assume that the magnetic field frequency $\omega$ equals to that of the sound wave. Writing down the real magnetization value $m$ as $m = m_{00} + m'$ we see from Eq. (2) that $m'$ is also proportional to $e^{-i\omega t}$ and the following relation holds

$$m' = \frac{m'_0}{1-i\omega\tau} - \frac{i\omega\tau}{1-i\omega\tau}\chi H. \qquad (3)$$

Following the general method of examining the influence of slow relaxation processes on the second viscosity [1, 2] we shall find the difference between the real and equilibrium pressure values $p - p_0$. The fluid pressure appears in the second equation of the system (1), that contains the second-viscosity coefficients $\xi_1$ and $\xi_2$. In addition, in case of superfluid solutions, one should also find the difference $\Phi - \Phi_0$ between the real and the equilibrium values of the thermodynamic potential $\Phi = (\mu - cz/\rho)$. This quantity enters the third equation of the system (1) that contains the second viscosities $\xi_3$ and $\xi_4$.

To find the pressure difference we choose the density $\rho$, the entropy $\sigma$ and the concentration $c$ as independent thermodynamic variables. The pressure is a function of these variables and of the magnetization. The derivative of $p$ with respect to one of the thermodynamic variables, say $\rho$, is of the form

$$\left(\frac{\partial p}{\partial \rho}\right)_{\sigma,c} = \left(\frac{\partial p}{\partial \rho}\right)_m + \left(\frac{\partial p}{\partial m}\right)_{\rho,\sigma,c}\left(\frac{\partial m}{\partial \rho}\right)_{\sigma,c}. \qquad (4)$$

According to the relation (3) we obtain

$$\left(\frac{\partial m}{\partial \rho}\right)_{\sigma,c} = \frac{1}{1-i\omega\tau}\left(\frac{\partial m'_0}{\partial \rho}\right)_{\sigma,c} - H\frac{\partial}{\partial \rho}\left(\frac{i\omega\tau}{1-i\omega\tau}\chi\right)_{\sigma,c} = \frac{1}{1-i\omega\tau}\left(\frac{\partial m_0}{\partial \rho}\right)_{\sigma,c} - H\frac{\partial}{\partial \rho}\left(\frac{i\omega\tau}{1-i\omega\tau}\chi\right)_{\sigma,c}.$$

In this relation we are taking into account that the relaxation time $\tau$ and the magnetic susceptibility $\chi$ may depend on thermodynamic variables. In addition we should take into account that the relaxation time of magnetization in quantum solutions is rather large. Depending on temperature and concentration it varies in the range between 10 and $10^3$ seconds (see [7]). This means that $\omega\tau \gg 1$, and thus the relation (4) can be rewritten as

$$\left(\frac{\partial p}{\partial \rho}\right)_{\sigma,c} = \frac{1}{-i\omega\tau}\left[\left(\frac{\partial p}{\partial \rho}\right)_{eq} - i\omega\tau\left(\frac{\partial p}{\partial \rho}\right)_m\right] + H\left(\frac{\partial p}{\partial m}\right)_{\rho,\sigma,c}\left(\frac{\partial \chi}{\partial \rho}\right)_{\sigma,c}, \qquad (5)$$

where the notation is introduced:

$$\left(\frac{\partial p}{\partial \rho}\right)_{eq} = \left(\frac{\partial p}{\partial \rho}\right)_m + \left(\frac{\partial p}{\partial m}\right)_{\rho,\sigma,c}\frac{\partial m_0}{\partial \rho}.$$

This expression is a derivative of the pressure in case the equilibrium state holds in the fluid. The derivatives of $p$ with respect to the other two thermodynamic variables, $\sigma$ and $c$, are of the similar form. If the entire pressure variation is

$$\left(\frac{\partial p}{\partial \rho}\right)_{\sigma,c} \delta\rho + \left(\frac{\partial p}{\partial \sigma}\right)_{\rho,c} \delta\sigma + \left(\frac{\partial p}{\partial c}\right)_{\rho,\sigma} \delta c,$$

and the equilibrium pressure variation equals to

$$\left(\frac{\partial p}{\partial \rho}\right)_{eq} \delta\rho + \left(\frac{\partial p}{\partial \sigma}\right)_{eq} \delta\sigma + \left(\frac{\partial p}{\partial c}\right)_{eq} \delta c,$$

then the difference $p - p_o$ for the case of the first sound is as follows

$$p - p_0 = \left\{ \left(\frac{\partial p}{\partial \rho}\right)_m - \left(\frac{\partial p}{\partial \rho}\right)_{eq} - \frac{\rho_s}{\rho_n}\frac{\varepsilon}{\rho} \left[ \sigma\left(\left(\frac{\partial p}{\partial \sigma}\right)_m - \left(\frac{\partial p}{\partial \sigma}\right)_{eq}\right) + c\left(\left(\frac{\partial p}{\partial c}\right)_m - \left(\frac{\partial p}{\partial c}\right)_{eq}\right) \right] \right.$$

$$\left. + H\left(\frac{\partial p}{\partial m}\right)_{\rho,\sigma,c} \left[ \frac{\partial \chi}{\partial \rho} - \frac{\rho_s}{\rho_n}\frac{\varepsilon}{\rho}\left(\sigma\frac{\partial \chi}{\partial \sigma} + c\frac{\partial \chi}{\partial c}\right) \right] \right\} \delta\rho = A_1 \, \delta\rho, \qquad (6)$$

where $A_1$ is the expression in the curly brackets, $\varepsilon = (c/\rho)(\partial\rho/\partial c)$. In this equality the relations between $\delta\rho$, $\delta\sigma$ and $\delta c$ in the first-sound wave are used:

$$\delta\sigma = -\frac{\rho_s}{\rho_n}\frac{\varepsilon}{\rho}\sigma\delta\rho; \qquad \delta c = -\frac{\rho_s}{\rho_n}\frac{\varepsilon}{\rho}c\delta\rho.$$

In a sound wave the quantity $\delta\rho$ is coupled with the velocities of the normal and superfluid flows with the continuity equation (the first equation of the system (1)). Since in the first-sound wave the normal and superfluid velocities are related via

$$\mathbf{v}_s = (\varepsilon + 1)\left(1 - \frac{\rho_s}{\rho_n}\varepsilon\right)^{-1} \mathbf{v_n},$$

the continuity equation yields

$$\delta\rho = \frac{\rho}{i\omega}\left(1 - \frac{\rho_s}{\rho_n}\varepsilon\right)^{-1} \text{div } \mathbf{v}_n.$$

Thus the difference $p - p_0$ between the real and equilibrium pressure values is

$$p - p_0 = A_1 \frac{\rho}{i\omega}\left(1 - \frac{\rho_s}{\rho_n}\varepsilon\right)^{-1} \text{div } \mathbf{v_n} \qquad (7)$$

Let us now consider the stress tensor $\sigma_{ik}$, which is equal in the linear approximation to the momentum flux density tensor $\Pi_{ik}$ with the opposite sign: $\sigma_{ik} = -(p - p_0)\delta_{ik} + \sigma'_{ik}$, where the equilibrium pressure $p_0$ is separated and a viscous part of the stress tensor $\sigma'_{ik}$ is added. Comparing Eq. (7) with the expression for $\sigma_{ik}$ one can see that in the presence of a relaxation process a term equal to the right-hand side of Eq. (7) with the opposite sign is added to $\sigma_{ik}$.

To connect this additional term with the second viscosity let us consider the right-hand side of the second equation in Eqs. (1). Using the relation (6) between the velocities $\mathbf{v}_s$ and $\mathbf{v}_n$ we get

$$\xi_1 \operatorname{div}(\mathbf{j} - \rho \mathbf{v}_n) + \xi_2 \operatorname{div} \mathbf{v}_n = \left[ \xi_1 \rho \frac{\rho_s}{\rho_n} \varepsilon \left(1 - \frac{\rho_s}{\rho_n} \varepsilon\right)^{-1} + \xi_2 \right] \operatorname{div} \mathbf{v}_n \qquad (8)$$

Comparison of this equality with the additional term in $\sigma_{ik}$ (cf. (7)) shows that in the presence of the magnetization relaxation movement the sum from Eqs. (1) containing the second viscosity coefficients $\xi_1$ and $\xi_2$ becomes of the form

$$\xi_1 \rho \frac{\rho_s}{\rho_n} \varepsilon + \xi_2 \left(1 - \frac{\rho_s}{\rho_n} \varepsilon\right) = A_1 \frac{\rho}{-i\omega}. \qquad (9)$$

In a similar way one can derive the expression for the second viscosity coefficients $\xi_3$ and $\xi_4$ (in this case it is convenient to choose $p$, $T$, $c$ as independent thermodynamic quantities):

$$\xi_3 \rho \frac{\rho_s}{\rho_n} \varepsilon + \xi_4 \left(1 - \frac{\rho_s}{\rho_n} \varepsilon\right) = B_1 \frac{\rho u_1^2}{-i\omega},$$

where $u_1$ is the first-sound velocity and $B_1$ equals to

$$B_1 = \left(\frac{\partial \Phi}{\partial p}\right)_m - \left(\frac{\partial \Phi}{\partial p}\right)_{eq} - \frac{\rho_s}{\rho_n} \frac{\varepsilon}{\rho u_1^2} \left\{ \overline{\sigma} \left(\frac{\partial \sigma}{\partial T}\right)^{-1} \left[\left(\frac{\partial \Phi}{\partial T}\right)_m - \left(\frac{\partial \Phi}{\partial T}\right)_{eq}\right] + c\left[\left(\frac{\partial \Phi}{\partial c}\right)_m - \left(\frac{\partial \Phi}{\partial c}\right)_{eq}\right]\right\} + H \left(\frac{\partial \Phi}{\partial m}\right)_{p,T,c} \left\{\frac{\partial \chi}{\partial p} - \frac{\rho_s}{\rho_n} \frac{\varepsilon}{\rho u_1^2} \left[\overline{\sigma}\left(\frac{\partial \sigma}{\partial T}\right)^{-1} \frac{\partial \chi}{\partial T} + c \frac{\partial \chi}{\partial c}\right]\right\}.$$

In obtaining the expression for $B_1$ the following relations between $\delta p$, $\delta T$ and $\delta c$ have been used:

$$\delta T = -\frac{\rho_s}{\rho_n} \frac{\varepsilon}{\rho u_1^2} \overline{\sigma} \left(\frac{\partial \sigma}{\partial T}\right)^{-1} \delta p; \qquad \delta c = -c \frac{\rho_s}{\rho_n} \frac{\varepsilon}{\rho u_1^2} \delta p.$$

Similar reasoning and calculations allow derivation of the expressions for the second viscosity coefficients in case of the second-sound wave propagation in the presence of an external oscillating magnetic field:

$$-\xi_1 \rho + \xi_2 (1+\varepsilon) = -A_2 (\rho/i\omega)\varepsilon, \qquad -\xi_3 \rho + \xi_4 (1+\varepsilon) = -B_2 (\rho u_2^2/i\omega)\varepsilon,$$

with $u_2$ being the second-sound velocity and $A_2$, $B_2$ equal to

$$A_2 = \left(\frac{\partial p}{\partial \rho}\right)_m - \left(\frac{\partial p}{\partial \rho}\right)_{eq} + \frac{\sigma}{\rho \varepsilon}\left[\left(\frac{\partial p}{\partial \sigma}\right)_m - \left(\frac{\partial p}{\partial \sigma}\right)_{eq}\right] + \left(\frac{\partial \rho}{\partial c}\right)^{-1}\left[\left(\frac{\partial p}{\partial c}\right)_m - \left(\frac{\partial p}{\partial c}\right)_{eq}\right] +$$

$$H\left(\frac{\partial p}{\partial m}\right)_{\rho,\sigma,c}\left[\frac{\partial \chi}{\partial \rho} + \frac{\sigma}{\rho \varepsilon}\frac{\partial \chi}{\partial \sigma} + \left(\frac{\partial \rho}{\partial c}\right)^{-1}\frac{\partial \chi}{\partial c}\right];$$

$$B_2 = \left(\frac{\partial \Phi}{\partial p}\right)_m - \left(\frac{\partial \Phi}{\partial p}\right)_{eq} + \frac{1}{\rho u_2^2 \varepsilon}\left\{\bar{\sigma}\left(\frac{\partial \sigma}{\partial T}\right)^{-1}\left[\left(\frac{\partial \Phi}{\partial T}\right)_m - \left(\frac{\partial \Phi}{\partial T}\right)_{eq}\right] + c\left[\left(\frac{\partial \Phi}{\partial c}\right)_m - \left(\frac{\partial \Phi}{\partial c}\right)_{eq}\right]\right\} +$$

$$H\left(\frac{\partial \Phi}{\partial m}\right)_{p,T,c}\left\{\frac{\partial \chi}{\partial p} + \frac{1}{\rho u_2^2 \varepsilon}\left[\bar{\sigma}\left(\frac{\partial \sigma}{\partial T}\right)^{-1}\frac{\partial \chi}{\partial T} + c\frac{\partial \chi}{\partial c}\right]\right\}.$$

We shall also examine the influence of the magnetic moment relaxation on the second viscosity in case of the fourth-sound wave propagation in superfluid solutions, that is for the case of sound-wave propagation in narrow capillaries. When helium is flowing in a narrow capillary the normal fluid component is locked by viscous forces and remains at rest with respect to the channel walls but sound waves can still propagate through motion of the superfluid component. Since the normal-flow velocity $\mathbf{v}_n = 0$ there remains only one second viscosity coefficient $\xi_3$ and the hydrodynamic equations take the form

$$\frac{\partial \rho}{\partial t} + \rho_s \operatorname{div} \mathbf{v}_s = 0,$$

$$\frac{\partial \mathbf{v}_s}{\partial t} + \nabla\left(\mu - c\frac{z}{\rho}\right) = \xi_3 \rho_s \nabla \operatorname{div} \mathbf{v}_s,$$

$$\frac{\partial (\rho c)}{\partial t} = 0,$$

$$\frac{\partial (\rho \sigma)}{\partial t} = 0. \tag{10}$$

The other second viscosity coefficients, $\xi_1, \xi_2$ and $\xi_4$, are not present in Eqs. (10). The coefficient $\xi_4$ disappears because of the lock of the normal velocity. The coefficients $\xi_1$ and $\xi_2$ disappear for the reason that now we cannot use the equation for the momentum flux density containing these coefficients because this equation does not include the dominant viscous force which reduces $\mathbf{v}_n$ to zero. On the other hand we don't need this equation to describe the forth-sound wave propagation because the number of variables has been reduced to six and we have six equations. Thus we have only one viscosity coefficient $\xi_3$ and calculations yield

$$\xi_3 = -B_4 \frac{1+\varepsilon}{i\omega} \frac{\rho}{\rho_s} u_4^2, \qquad \omega \gg 1,$$

where $u_4$ is the fourth sound velocity and

$$B_4 = \left(\frac{\partial \Phi}{\partial p}\right)_m - \left(\frac{\partial \Phi}{\partial p}\right)_{eq} + \frac{\rho_s}{\rho^2 u_4^2 (1+\varepsilon)} \left\{ \overline{\sigma} \left(\frac{\partial \sigma}{\partial T}\right)^{-1} \left[ \left(\frac{\partial \Phi}{\partial T}\right)_m - \left(\frac{\partial \Phi}{\partial T}\right)_{eq} \right] + c \left[ \left(\frac{\partial \Phi}{\partial c}\right)_m - \left(\frac{\partial \Phi}{\partial c}\right)_{eq} \right] \right\} +$$
$$H \left(\frac{\partial \Phi}{\partial m}\right)_{p,T,c} \left\{ \frac{\partial \chi}{\partial p} - \frac{\rho_s}{\rho^2 u_4^2 (1+\varepsilon)} \left[ \overline{\sigma} \left(\frac{\partial \sigma}{\partial T}\right)^{-1} \frac{\partial \chi}{\partial T} + c \frac{\partial \chi}{\partial c} \right] \right\}.$$

Now let us dwell on the signs of the expressions $A_{1,2}$ and $B_{1,2,4}$. Consider $A_1$ (the first sound). The equality (9) is valid for $\omega\tau \gg 1$. For arbitrary $\omega\tau$ values Eq. (9) becomes as follows

$$\xi_1 \rho (\rho_s / \rho_n) \varepsilon + \xi_2 [1 - (\rho_s / \rho_n) \varepsilon] = A_1 \rho \tau / (1 - i\omega\tau) \qquad (11)$$

It can be seen from (11) that for a relaxation process with $\omega\tau \ll 1$ and in the absence of an external magnetic field the quantity $A_1$ is positive (at least for $(\rho_s / \rho_n)\varepsilon < 1$ which is valid for dilute solutions at not very low temperature):

$$\left(\frac{\partial p}{\partial \rho}\right)_m - \left(\frac{\partial p}{\partial \rho}\right)_{eq} - \frac{\rho_s}{\rho_n} \frac{\varepsilon}{\rho} \left[ \sigma \left( \left(\frac{\partial p}{\partial \sigma}\right)_m - \left(\frac{\partial p}{\partial \sigma}\right)_{eq} \right) + c \left( \left(\frac{\partial p}{\partial c}\right)_m - \left(\frac{\partial p}{\partial c}\right)_{eq} \right) \right] > 0$$

In the presence of an external oscillating magnetic field the relaxation process will reduce the second viscosity if the part of $A_1$ containing magnetic field is negative. This depends on the values of the thermodynamic quantities, on the derivatives of the pressure and magnetic susceptibility and also on the phase $\varphi$ of the magnetic field ($H = |H| e^{i\varphi}$). The phase $\varphi$ can be chosen equal either to zero or $\pi$. The absence of relevant experimental data hinders performing numerical estimates, but anyway the relaxation motion of the magnetization in a superfluid solution subjected to an external oscillating magnetic field could lead to decreasing the second viscosity or, more precisely, the sum $\xi_1 \rho(\rho_s / \rho_n)\varepsilon + \xi_2 [1 - (\rho_s / \rho_n)\varepsilon]$ provided

$$|H| e^{i\varphi} \left(\frac{\partial p}{\partial m}\right)_{\rho,\sigma,c} \left[ \frac{\partial \chi}{\partial \rho} - \frac{\rho_s}{\rho_n} \frac{\varepsilon}{\rho} \left( \sigma \frac{\partial \chi}{\partial \sigma} + c \frac{\partial \chi}{\partial c} \right) \right] < 0$$

Similar reasoning is valid for $A_2$ and $B_{1,2,4}$.